\documentclass[aps, twocolumn,prx, showpacs, longbibliography, preprintnumbers,amsmath,amssymb,superscriptaddress]{revtex4-1}
\usepackage{amsmath,amsfonts,amssymb,graphics,graphicx,epsfig,color,times,indentfirst,layout,hyperref,subfigure,multirow,bm}
\usepackage{hyperref, soul}
\usepackage{ulem}
\hypersetup{linktocpage,,colorlinks=true}
\usepackage{threeparttable}
\usepackage{dcolumn}
\usepackage{bm}
\usepackage{color}

\begin{document}

    \title{Disorder-induced topology in quench dynamics}
    \author{Hsiu-Chuan Hsu}\email{hcjhsu@nccu.edu.tw}
    \affiliation{Graduate Institute of Applied Physics, National Chengchi University, Taipei 11605, Taiwan}
     \affiliation{Department of Computer Science, National Chengchi University, Taipei 11605, Taiwan}
    \author{Pok-Man Chiu} 
    \affiliation{Department of Physics, National Tsing Hua University, Hsinchu 30013, Taiwan}
    \author{Po-Yao Chang}\email{pychang@phys.nthu.edu.tw}
     \affiliation{Department of Physics, National Tsing Hua University, Hsinchu 30013, Taiwan}
    
    \date{\today}
\begin{abstract}
	We study the effect of strong disorder on topology and entanglement in quench dynamics. 
        Although disorder-induced topological phases have been well studied in equilibrium, the disorder-induced topology in quench dynamics has not been
        explored. In this work, we predict a disorder-induced topology of post-quench states
         characterized by the quantized dynamical Chern number 
	and the crossings in the entanglement spectrum in $(1+1)$ dimensions. 
	The dynamical Chern number undergoes transitions from zero to unity, and back to zero when increasing the disorder strength. 
	The boundaries between different dynamical Chern numbers are determined by  
         delocalized critical points in the post-quench Hamiltonian with the strong disorder. 
         An experimental realization in quantum walks is discussed.        

\end{abstract}
\maketitle
\section{Introduction}
Topological phases of matter out-of-equilibrium and their phase transitions
have attracted much theoretical and experimental interest. 
Their topological and non-equilibrium features have been demonstrated in various systems including
ultracold-atomic gases \cite{Foster2014,Plekhanov2017,Cooper2019,Salerno2019}, 
quantum optics \cite{Rechtsman2013,Wang2019_1,Ozawa2019},
superconducting qubits \cite{Kyriienko2018, Malz2020}, and condensed matter systems \cite{Kitagawa2011,Ezawa2013,Kundu2014,Gul2015,Farrell2015, Takasan2017,Owerre2018,Lubatsch2019,Oka2019}.
Among these, the topological Floquet systems have been widely studied \cite{Kitagawa2010, Lindner2011, Jiang2011}. 
These systems exhibit protected boundary states which are robust in the presence of disorder. 
More recently, topological phases in dynamical quench systems are proposed~\cite{Yang2018, Gong2018, Chang2018, Zhu2020, Hu2020}.
For example, for a trivial state under a sudden quench by the Su-Schrieffer-Heeger (SSH) model, 
the topology of the post-quench state is characterized by
the dynamical Chern numbers \cite{Yang2018, Chang2018}, 
the quantization of which describes a skyrmion texture of the post-quench pseudospin in the momentum-time space \cite{Wang2019}. The topology in quench dynamics has been shown experimentally in photonic quantum walks \cite{Cardano2017,Wang2019,Xu2019} and superconducting qubits \cite{Flurin2017,Guo2019}. 
Moreover, the entanglement spectrum provides an additional probe of the topology.
The robustness of crossings in the entanglement spectrum of the post-quench states
indicates the nontrivial topology in quench dynamics \cite{Gong2018, Chang2018}.

Besides the topological structures that emerge in quench dynamics, 
non-trivial topology can arise from disordered systems in equilibrium.
In the strong disorder regime, an unexpected topological phase with extensive boundary states
is stabilized by the strong disorder.
This phase is termed the ``topological Anderson insulator" \cite{Li2009,Jiang2009a,Groth2009a, Guo2010a, Hsu2020} and the transition
between trivial and non-trivial phases is described by the delocalization criticality~\cite{Shem2014}.
A generalization of the topological Anderson insulator to Floquet systems is proposed \cite{Titum2015,Titum2016,Titum2017,Liu2020}.
The strong disorder drives trivial Floquet systems into topological phases that host
chiral edge modes coexisting with the localized bulk states in two-dimensional lattices. The transition also links to delocalization \cite{Wauters2019}. 
In constrast, quench Anderson disorder was studied theoretically in simple lattice models \cite{Rahmani2018,Lundgren2019}.
It has been shown that in the strong disorder regime, where the Anderson localization sets in, there is no sharp transition in the quench dynamics~\cite{Rahmani2018}.

Although there are extensive studies in disorder-induced topology in Floquet systems,
the effect of disorder on topology in quench dynamics is less discussed.
It is shown that the crossings in the entanglement spectrum
are robust against weak disorder and interactions \cite{Gong2018}.
However, it has not been known whether disorder could induce topology in quench dynamics. 

In this work, we demonstrate the strong disorder-induced topology in quench dynamics. 
We consider a quench protocol described by a trivial initial state (a fully pseudospin-polarized state) under a sudden quench by 
the SSH Hamiltonian in the presence of strong disorder.
The topology of the post-quench state is characterized by the dynamical Chern number which is zero/unity when the SSH model is trivial/non-trivial.
We start at the clean limit where the post-quench state is trivial. 
When the disorder strength is above the critical value, the post-quench state has a
quantized dynamical Chern number.
The entanglement spectrum of the post-quench states
shows robust crossings which indicate the disorder-induced topology in quench dynamics.
The post-quench SSH Hamiltonian in this strong disorder regime has a 
disorder-induced winding number. The phase boundaries coincide with the transitions between vanishing and 
 quantized dynamical Chern numbers.  
Our results demonstrate that the disorder-induced topology in quench dynamics in $(1+1)$ dimensions 
is directly related to the topological Anderson insulator.



\section{The post-quench Hamiltonian}\label{sec:protocol}
We consider an eigenstate $|\Psi_0\rangle$ of a pre-quench Hamiltonian $H_{\rm pre}$ 
at $t=0$ under a sudden quench by a post-quench Hamiltonian $H_{\rm post}$, and 
the post-quench state is  $|\Psi(t)\rangle = \exp[- i H_{\rm post} t] | \Psi_0\rangle$.
We consider $H_{\rm post} =H_0+H_U$, with
\begin{align}
H_0&=\sum_{x=1}^{N_x} J_0 c^{\dagger}_{x,a}c_{x,b} + J_1 c^{\dagger}_{x+1,a}c_{x,b}+{\rm h.c.}, \notag\\
H_U&=\sum_{x=1}^{N_x}U_{1x}c^{\dagger}_{x,a}c_{x,b}+U_{2x}c^{\dagger}_{x,a}c_{x+1,b}+{\rm h.c.},
\label{disorder}
\end{align}
where $H_0$ is the SSH Hamiltonian and $H_U$ is the time-reversal and particle-hole symmetry preserving disorder.
Here $x$ is the label of the unit-cell, $N_x$ is the total number of the unit-cell.
$c^{\dagger}_{xa(b)}, c_{xa(b)}$ are the creation and annihilation
operators on sublattices $a, b$ on the $x$-th unit-cell. $J_{0(1)}$ denotes the
intracell(intercell) coupling, and $U_{1(2)x}$ is the random intracell(intercell) coupling strength given by the random number in
the uniform distribution $\left[-W_{1(2)}/2,W_{1(2)}/2\right]$.
We choose the disorder strengths $W_1=2W_2=W_0$.
The post-quench Hamiltonian $H_{\rm post}$ has the time-reversal symmetry $T: c_{xa(b)} \to c_{xa(b)}$, $i \to -i$,
and the particle-hole symmetry $C: c_{xa(b)} \to c_{xb(a)}$, $i \to -i$. I.e., it belongs to the BDI symmetry class, $T^2=C^2=1$. 

The topology of the post-quench Hamiltonian  $H_{\rm post}$ in the presence of strong disorder is characterized by the winding number $W$
and the phase diagram is shown in Fig. \ref{clean}(a). 
In the clean limit with 
$(J_0/J_1,W_0/J_1) = (1.1,0)$, the winding number is zero.
When the disorder strength increases, the winding number becomes unity when  $W_0/J_1 \gtrsim 1.7$
and is back to zero when $W_0/J_1\gtrsim 3.6$ [the white dash in Fig. \ref{clean}(a)].
This behavior demonstrates the disorder-induced quantized winding number in the post-quench Hamiltonian
and is referred to  
a topological Anderson insulator \cite{Meier929}. 
The phase boundaries of the trivial and the topological Anderson insulating phases are obtained by the divergence of the localization length $\lambda$~\cite{Shem2014} [see App.~\ref{App:Loc}],
\begin{align}
\frac{1}{\lambda}=\bigg|\ln\left[
\frac{\big|2J_1+W_1 \big|^{\frac{J_1}{W_1}+\frac{1}{2}}\big|2J_0-W_2 \big|^{\frac{J_0}{W_2}-\frac{1}{2}}}{\big|2J_1-W_1 \big|^{\frac{J_1}{W_1}-\frac{1}{2}}\big|2J_0+W_2 \big|^{\frac{J_0}{W_2}+\frac{1}{2}}}
\right]
\bigg|.
\end{align}

\section{The quench protocol}
In the clean limit, the post-quench
Hamiltonian is diagonalized in the momentum space $H_{\rm post} = \sum_k \psi^\dagger_k \mathcal{H}_{\rm post}(k) \psi_k$
with $ \psi_k = (c_{ka},c_{kb})^{\rm T}$ with eigenenergies $\pm |E(k)|$. Since each single-particle state does not interact with each other, 
the single-particle state evolves individually $|\psi(k,t ) \rangle= e^{- i \mathcal{H}_{\rm post}(k) t} |\psi_0(k) \rangle$, where $|\psi_0(k) \rangle$ is the single-particle ground state of the pre-quenched single-particle Hamiltonian $\mathcal{H}_{\rm pre}(k)$.
For each individual post-quench single-particle state, the period of the dynamics is $T_k=2\pi/|E(k)|$.
The set of single-particle states $|\psi(k,t ) \rangle$ have a corresponding 
momentum-time manifold $k \in [0, 2\pi]$, $t_k \in [0, T_k]$ which is a momentum-time torus.
This torus is distorted because different $k$ has different circumference $T_k$.
Since the deformation of the distorted torus to a ordinary torus (same circumference) does not change the topology, 
one can rescale the period of the dynamics to be $T_k=2\pi$
The rescaling of the period is equivalent to flattening the post-quench Hamiltonian, 
 $\mathcal{H}^{F} (k) = \mathcal{H}_{\rm post} (k)/|E_{(k)}|$. 
 We focus on the flattened Hamiltonian 
 which allows us to construct the effective Hamiltonian $\mathcal{H}_{\rm eff} (k ,t) = e^{-i \mathcal{H}^F(k) t} \mathcal{H}_{\rm pre} (k)e^{i \mathcal{H}^F(k) t}$
 for analyzing the topological property of the post-quench dynamics~[see App. \ref{App:Symm}].

 \subsection{Different pre-quench Hamiltonians} 
 The post-quench state has two inputs, the pre-quench Hamiltonian $\mathcal{H}_{\rm pre}(k)$ and the post-quench Hamiltonian
$\mathcal{H}_{0}(k)$.
If the pre-quench and the post-quench Hamiltonians are in the same symmetry class (BDI), the topology of
the post-quench state is characterized by the dynamical Chern number in the half of the Brillouin zone (BZ), $k \in [0, \pi]$ and $t \in \left[0,\pi\right]$ \cite{Yang2018}.
However, the dynamical Chern number is vanishing in the full BZ, $k\in [0,2\pi]$ and $t \in \left[0,\pi\right]$.
To study the disorder-induced topology in quench dynamics, the real-space formalism is needed and requires
the information of the full BZ.
Since the dynamical Chern number vanishes in the full BZ and $t \in \left[0,\pi\right]$, no disorder-induced topology can happen in this quench protocol.
On the other hand, if the pre-quench Hamiltonian $\mathcal{H}_{\rm pre}(k)=-\sigma_z$ which is not in the same symmetry class
as the post-quench Hamiltonian, the dynamical Chern number is quantized in the full BZ, $t \in \left[0,\pi/2 \right]$ \cite{Chang2018} [see App.~\ref{App:Symm}]. 
This pre-quench Hamiltonian allows us to formulate the dynamical Chern number in real-space 
and study the disorder-induced topology.

In this case, 
the single-particle state  is fully pseudospin polarized and
the real-space expression is $|\psi_i\rangle=   \left( 1, \: 0\right)^{\rm T}\otimes|i\rangle$,
where $ \left( 1, \: 0\right)^{\rm T}$ denotes one particle at the sublattice $a$, $|i\rangle = (0,\cdots, 1, \cdots, 0)^{\rm T}$ denotes the only non-vanishing $i$-th element with $i$ being the site label
$i =1\dots N_x$. 
The post-quench Hamiltonian in the presence of the disorder can be flattened 
by using the projectors, $\mathcal{H}^F=|\psi_{+}\rangle\langle\psi_{+}|-|\psi_{-}\rangle\langle\psi_{-}|$,
where $|\psi_{\pm}\rangle$ are the eigenstates of $\mathcal{H}$ with positive/negative energies. 

\subsection{Berry phase and dynamical Chern number}\label{sec:berry}
\begin{figure}[]
\includegraphics[width=0.5\textwidth]{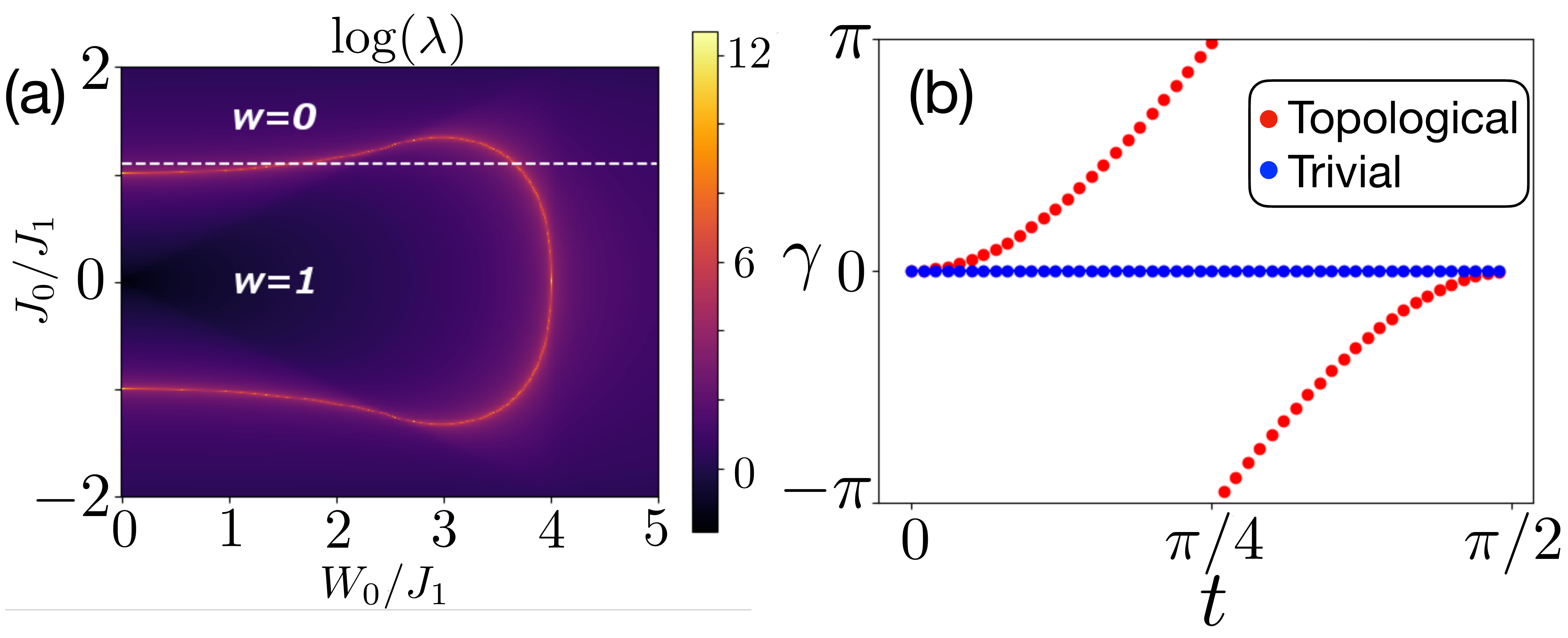}
	\caption{(a) The phase diagram of the post-quench Hamiltonian $H=H_0+H_U$.
	          The white dashed line denotes $J_0=1.1$. 
		(b) The time-dependent Berry phase in the clean limit. The blue dots are for the trivial post-quench state ($J_0/J_1=1.1$).
	The red dots are for the topological post-quench state with a Berry phase flow from $t=0$ to $\pi/2$ ($J_0/J_1=0.5$).
	}
	\label{clean}
\end{figure}

To determine the dynamical Chern number in the real space, we compute the Berry phase with the twisted boundary condition~\cite{Niu1985a, Qi2006b} by the overlap matrix~\cite{Gresch2017,Kuno2019,Bonini2020}. 
 The overlap matrix at a $t$ is defined as
$M_{ij}^{\ell}(t)=\langle \psi^{\theta_\ell}_i(t)|\psi^{\theta_{\ell+1}}_j(t)\rangle$,
where $|\psi_{i}^{\theta_{\ell}}(t)\rangle= \exp [-i {\mathcal{H}^{\theta_\ell}_{\rm post}} t]|\psi_{i}\rangle$, $i$ is the index of the single-particle state, and $\mathcal{H}^{\theta_\ell}_{\rm post}$ is the flattened post-quench Hamiltonian with twisted boundary phase 
${\theta_\ell}=\frac{2\pi\ell}{L}$
~\cite{Kuno2019}, where $L$ is the number of mesh points and $l=1,\cdots,L$. The Berry phase is given by 
$\gamma(t)= {\rm Im } \left [ \ln \det \prod_{\ell=1}^L M^{\ell}(t) \right]$. 
The Berry phase as a function of $t$ has no jump when the post-quench state is trivial [Fig.\ref{clean}(b) blue dots].
In contrast, when the post-quench state is topological,
 the Berry phase flow has $2 \pi$ jumps at  $t=\pi/4$ as shown by the red dots in Fig.\ref{clean}(b).
The Wannier center flow also shows similar behavior which we demonstrate in the App. \ref{App:Wannier}.

\begin{figure}[]
\includegraphics[scale=0.5]{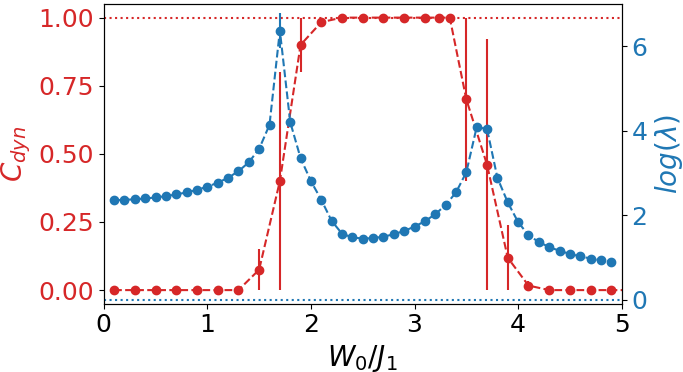}
	\caption{The disorder-average  dynamical Chern number and the localization length for the post-quench Hamiltonian. 
	The error bar is the standard deviation. 
	There are more than $20$ disorder realizations for each data point. The parameters are $J_0/J_1=1.1 , N_x=400$. 
	}
	\label{meanwind}
\end{figure}

The dynamical Chern number is obtained by integrating the time derivative of the Berry phase
$C_{\rm dyn}=\frac{1}{2\pi}\int^{\pi/2}_0 dt \frac{\partial\gamma(t)}{\partial t}$.
Since $t=\pi/2$ is the time taken for the pseudospin to precess from the north pole to the south pole, 
the integration is equivalent to counting the numbers of the pseudospin $\hat{n}_i(t)=\langle \psi_i(t)|\vec{\sigma}|\psi_i(t)\rangle$ wrapping around the entire Bloch sphere~\cite{Chang2018}.
The disorder-induced dynamical Chern number is shown in the red dots in Fig.~\ref{meanwind}(a).  
 In the weak disorder limit, $W_0/J_1 \lesssim 2.2 $ and $J_0/J_1=1.1$,
 the dynamical Chern number vanishes.
While increasing the disorder strength $W_0$, 
the dynamical Chern number is quantized with negligible fluctuations in the region $2.2 \lesssim W_0 \lesssim 3.2 $.
This behavior demonstrates that the disorder drives the trivial post-quench state to be topological, and
we refer it to the disorder-induced topology in quench dynamics.

The phase boundaries of the zero and unity dynamical Chern numbers 
coincide with the phase boundaries of the post-quench Hamiltonian obtained from the divergence of the localization length [the white dashed line in Fig. \ref{clean} (a)
and the blue dots in Fig.~\ref{meanwind}].
It was demonstrated that in the clean limit, 
the topology of the quench dynamics is related to that of the post-quench Hamiltonian~\cite{Chang2018, Gong2018}.
Here, we observe that the relation is still held for the disorder-induced topology.

\subsection{Entanglement spectrum}
The entanglement spectrum provides the additional information of the topology induced by disorder in quench dynamics. 
It is shown that the crossings in the entanglement spectrum reveal the topological properties in both the equilibrium systems \cite{Li2008, Pollmann2010, Fidkowski2010, Turner2010, Peschel2011, Hughes2011,Chang2014} and out-of-equilibrium systems~\cite{Gong2018,Chang2018,McGinley2018,Pastori2020}.
The presence/absence of the robust crossings in the entanglement spectrum indicates the post-quench state is topological/trivial. To compute the entanglement properties,
 the system is bipartite spatially  into $A$ and $B$ subsystems, where the post-quench many-body state is expressed as $|\Psi(t)\rangle = \sum_{i,j} C_{ij}(t) |A_i\rangle|B_j\rangle$
 with $|A(B)_i\rangle$ being the local basis in subsystem $A(B)$. We can compute the reduced density matrix $\rho_A (t)= {\rm Tr}_B |\Psi(t)\rangle \langle \Psi(t)| = \frac{1}{N} e^{-H_A(t)}$,
 where $H_A(t)$ is referred to the entanglement Hamiltonian, $N$ is the normalization constant, and the spectrum of $H_A(t)$ is the entanglement spectrum.
 
 In free-fermion systems, the eigenvalues of the reduced density matrix can be obtained from the correlation matrix 
 $C_{\bf x, x'}(t) = \langle\Psi(t)| c^\dagger_{\bf x} c_{\bf x'} |\Psi(t) \rangle= \sum_i | \psi_i({\bf x'},t)  \rangle \langle \psi_i({\bf x},t)  |$,
 where $| \psi_i({\bf x},t)  \rangle$ is the postquench single-particle state~[see App. \ref{App:CA}]. The spectrum $\xi(t)$ of the correlation matrix $C_{\bf x, x'}(t) $ with $x, x'$ being restricted in $A$ is related to the entanglement spectrum $\epsilon(t)$ by $\xi(t) = 1/(1+e^{\epsilon(t)})$ \cite{Peschel2003}. For simplicity, we refer $\xi(t)$ to the entanglement spectrum.

\begin{figure}[]
	\includegraphics[width=0.5\textwidth]{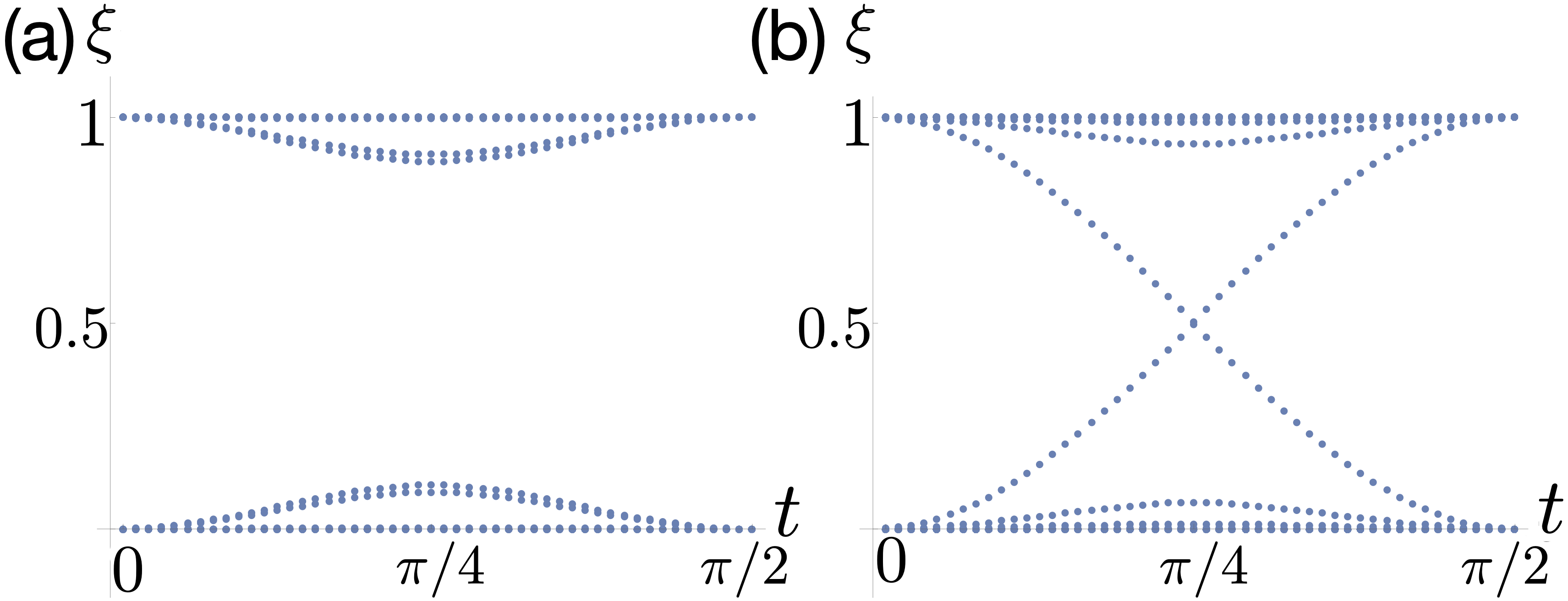}
	\caption{The entanglement spectrum of the postquench state with the bipartition $l_A=l_B=N_x/2$, where $l_{A(B)}$ 
	is the length of the subsystem $A(B)$ and $N_x$ is the length of the total system. The parameters are $J_0/J_1=1.1$. (a) $W_0=0$ (clean limit). (b) $W_0=3$. There are $100$ disorder realizations for each data point.
	}
	\label{ES}
\end{figure}

In the clean limit at $J_0/J_1=1.1$ [Fig. \ref{ES}(a)], 
the post-quench state is trivial and no crossings in the entanglement spectrum $\xi(t)$.
When the disorder strength is above the critical values, the entanglement spectrum  $\xi(t)$ of the post-quench state shows a crossing at $t=\pi/4$ [Fig. \ref{ES}(b)].
The existence of the crossings in the entanglement spectrum agrees with the non-vanishing dynamical Chern number of the post-quench state.
We demonstrate the non-vanishing dynamical Chern number and the crossings in the entanglement spectrum for other parameters in App. \ref{App:other}.

\section{Experimental realization}
Discrete-time quantum walks are great platforms for simulating the topological phases of matter~\cite{Kitagawa2011b,Cardano2017,Wang2018_2},
quantum quenches~\cite{Wang2019, Xu2019}, and disorder phenomena~\cite{Obuse2011,Zeng2017,Kumar2018}. 
Following Ref.~\cite{Wang2019}, the discrete-time evolution operator for a one-dimensional lattice
with single photons can be engineered by the cascaded half-wave plates and beam displacers. 
The Hilbert space is spanned by the polarization states $\{ |P_+\rangle, |P_-\rangle \}$ and the position state $|x\rangle$ with
$x \in \mathbb{Z}$.
The corresponding evolution operator for each time step is $U= R(\phi_1/2) S R(\phi_2)SR(\phi_1/2)$,
where $R(\phi)$ rotate the polarization by $\phi$ with respect to $y$-axis, and $S$ is the shift operator
$S=\sum_x |x-1 \rangle \langle x | \otimes |P_+\rangle \langle P_+|+|x+1 \rangle \langle x | \otimes |P_-\rangle \langle P_-|$.
The polarization angle $\phi_{1(2)}(x)$ is spatially dependent and disorder can be introduced by choosing different $\phi_{1(2)}(x)$ for different position $x$.
\begin{figure}[]
	\includegraphics[width=0.5\textwidth]{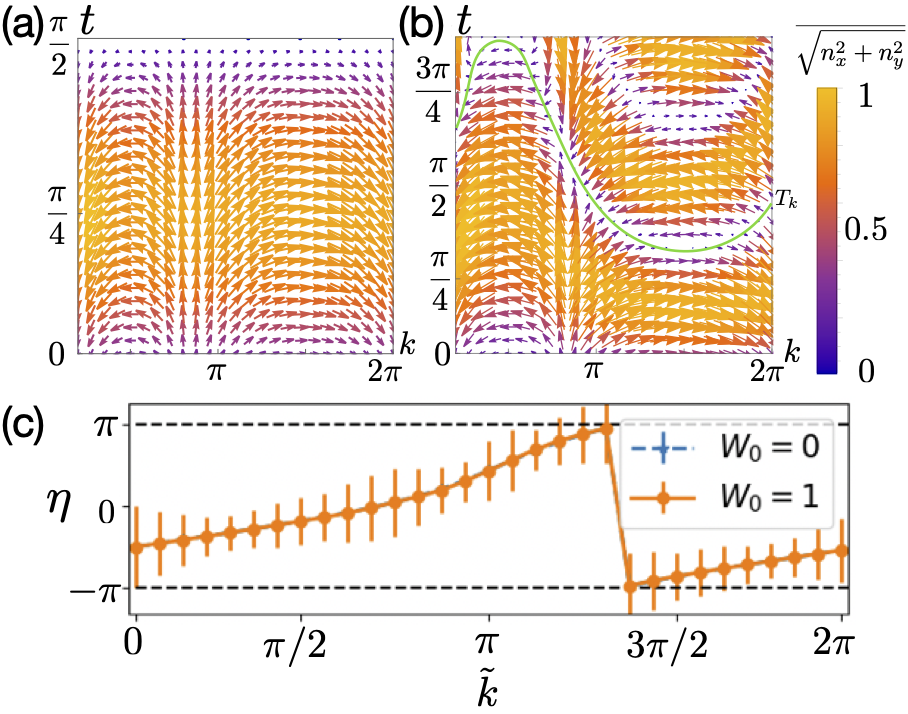}
	\caption{
	The post-quench psuedospin texture in the momentum-time space without disorder with $J_0/J_1=0.5$ for
          (a)flattened Hamiltonian,  
	 (b) non-flattened Hamiltonian. The post-quench psuedospin forms the Skyrmion texture in the momentum-time domain, $k\in [0, 2\pi] $,
	   $t\in [0, \pi/2]$ in (a) and $t \in [0, T_k]$  in (b) where $T_k= \pi/(2E(k))$ is shown by the green line.
         (c) The long-time average of $\eta$ for the non-flattened Hamiltonian with  $J_0/J_1=0.5$. Two curves almost overlap. The error bars are the standard deviation for $400$ disorder realization. $N_x=31$.}
	\label{skry}
\end{figure}

In a translation-invariant case, the unitary operator can be diagonalized in the momentum space and the effective Hamiltonian describing the pre/post-quench system
has the form $\mathcal{H}_{\rm eff}(k) = - i \ln U(k)$.
It is shown that this quantum walk protocol \cite{Wang2019} can simulate a sudden quench between $\mathcal{H}^{\rm i}_{\rm eff}(k)$
and $\mathcal{H}^{\rm f}_{\rm eff}(k)$ of the SSH model. Here $\mathcal{H}^{\rm i}_{\rm eff}(k)$ and $\mathcal{H}^{\rm f}_{\rm eff}(k)$ are referred to the pre-quench and the post-quench Hamiltonians.
The topology of the post-quench state can be extrapolated from the post-quench pseudospin ${\bf n}(k, t)={\rm Tr} [\rho(k,t) \boldsymbol{\sigma}]$ with $\rho(k,t) = |\psi_k(t)\rangle \langle \psi_k(t)|$.
The post-quench pseudospin forms the Skyrmion texture in the momentum-time domain when the post-quench state
has non-vanishing dynamical Chern number [Fig.  \ref{skry}(a)]. The Skyrmion texture can be understood as 
the pseudospin pointing along the $+(-) z$ direction at $t=0 (\pi/2)$ and rotating clockwise as a function of $k$ on the $x-y$ plane.
In the experimental setup, the Hamiltonian is non-flatten and the period of dynamics of each momentum is $T_k= \pi/(2E(k))$.
Nevertheless, the Skyrmion texture of the pseudospin can be observed in the momentum-time domain $k\in [0, 2\pi]$, $t_k \in [0,T_k ]$ [Fig.  \ref{skry}(b)]
and was measured experimentally in the quantum walk setup \cite{Wang2019}.

In the presence of disorder, the momentum is no longer a good quantum number and the momentum-dependent period is not well-defined. For the non-flattened post-quench Hamiltoinan, we propose to measure the long-time average of the pseudospins $\overline{\langle \sigma_i \rangle_T}=\frac{1}{T}\int_0^{T} dt \overline{\langle \sigma_i\rangle} $, where $\langle \sigma_i\rangle={\rm Tr}\left[\rho'(\tilde{k},t)\sigma_i\right]$ and 
\begin{align}
\rho'(\tilde{k},t) = \overline{ \frac{1}{2} \sum_{i=0}^3 \sum_{x_1,x_2} e^{- i \tilde{k}(x_1-x_2)} \langle \psi_{x_1}(t) | \sigma_i | \psi_{x_2}(t)\rangle  \sigma_i }.
\end{align}
$\rho'(\tilde{k},t)$ is the disorder-averaged density matrix in the pseudomomentum-time space, where $\overline{\cdots}$ denotes the disorder average. 
Here $\tilde{k}$ is referred to the pseudomomentum, which indicates that the momentum is no longer a good quantum number in disordered systems.
	Since the $x,y-$ components of the Skyrmion texture shows a $2 \pi$ winding as a function of the pseudomomemtum $\tilde{k}$,
	one can monitor the in-plane pseudospin texture to detect the nontrivial topology by defining 
	\begin{eqnarray}
	\eta={\rm Im }\log\left[\overline{\langle\sigma_x\rangle}_T+
	i\overline{\langle\sigma_y\rangle}_T\right].
	\end{eqnarray}
	If the the post-quench state is topological, $\eta$ shows a $2\pi$ difference in
	$\tilde{k}=0$ to $2\pi$. 

\begin{figure}[]
	\includegraphics[width=0.5\textwidth]{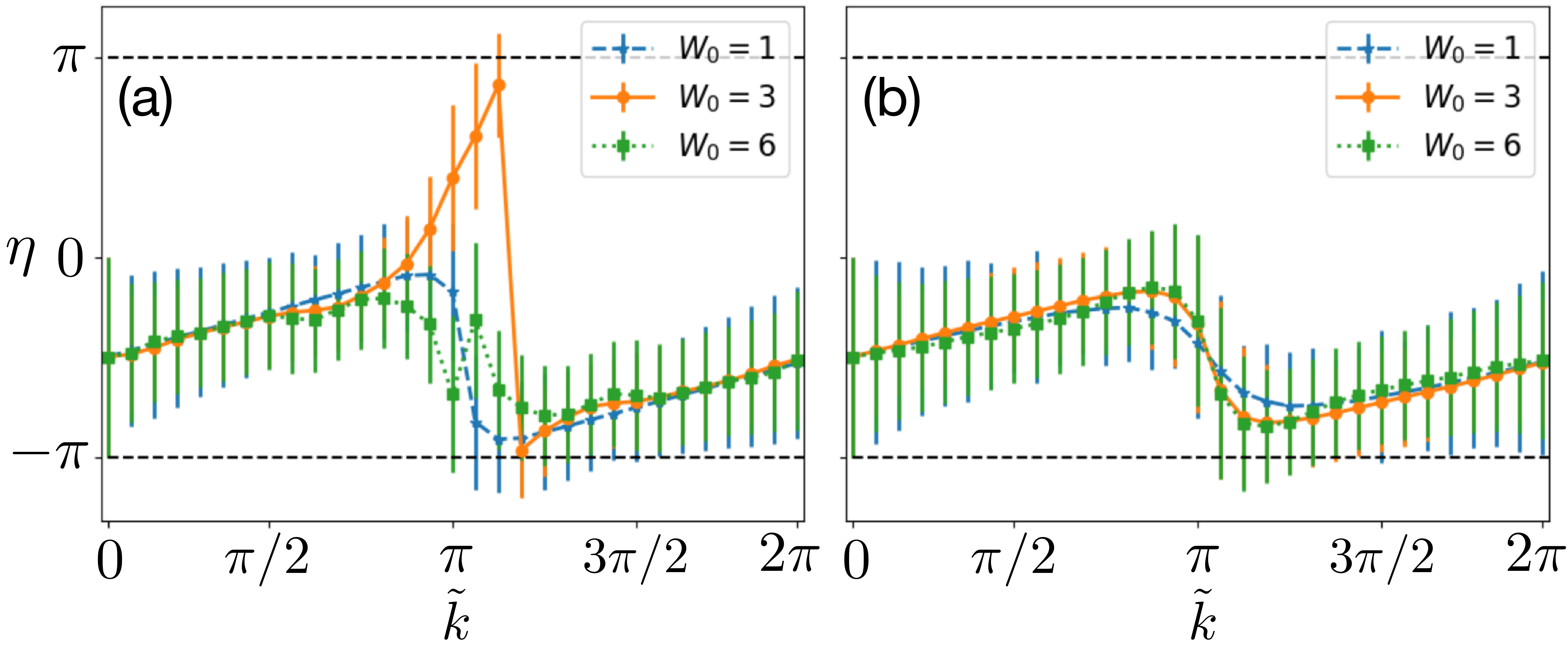}
	\caption{The long-time avarage of $\eta$ for the post-quench psuedospin in the pseudomomentum space given by non-flattened Hamiltonian with
		(a) $\frac{J_0}{J_1}=1.1$,
		and (b) $\frac{J_0}{J_1}=1.5$.
		The error bars are the standard deviations for $400$ disorder realizations. $N_x=31$.}
	\label{skry2}
\end{figure}

	We numerically show that $\eta$ can detect the topology of the post-quench state in Fig. \ref{skry} (c) and \ref{skry2}.
	The time taken for the average is $T=\pi/E_{min}$, where $E_{min}$ is the minimum absolute eigenenergy of the post-quench Hamiltonian in the clean limit. 
	 This average time $T$ is the largest time-scale in the system.
	First, we demonstrate the topology of post-quench state is robust in the weak disorder region. As shown in Fig. \ref{skry} (c),
	the in-plane pseudospin angle $\eta$ exhibits a $2\pi$ winding in the clean limit $W_0/J_1=0$ and the weak disorder region $W_0/J_1=1$ for the 
	parameter $J_0/J_1=0.5$.
	Next, we consider the disorder-induced topology for the parameter $J_0/J_1=1.1$. As we demonstrate previously,
	 the post-quench state is topological for the disorder strength $1.7\lesssim W_0/J_1\lesssim 3.6$. 
	 As shown in Fig. \ref{skry2} (a), $\eta$ does not has a $2\pi$ winding at $W_0/J_1=1$ and $W_0/J_1=6$, but exhibits a $2\pi$ winding at $W_0/J_1=3$, reflecting the disorder-induced topology. In contrast, for the parameter $J_0/J_1=1.5$ which does not exhibit the disorder-induced topology,
$\eta$ does not show a $2\pi$ winding with different strong disorder strengths as shown in Fig. \ref{skry2} (b).


\section{Conclusion}\label{conclusion}
 We predict the disorder-induced topology in quench dynamics in (1+1) dimensions. 
 The topology is characterized by the dynamical Chern number and crossings in the entanglement spectrum. 
 We show the boundaries between trivial and nontrivial post-quench states
 are identified by delocalized critical points in the post-quench Hamiltonian. 
The quantized dynamical Chern number in $(1+1)$ dimensions corresponds to the 
winding number of the one-dimensional topological Anderson insulating phase of the SSH model. 
 Finally, we propose this phenomenon can be realized in quantum walk experiments.

\begin{acknowledgments}
The authors thank Ching-Hao Chang and  Chao-Cheng Kaun for hosting the workshop of quantum materials at  
Research Center for Applied Sciences, Academia Sinica, where the work was partially initiated.
H.C.H. was supported by the Ministry of Science and Technology (MOST) in Taiwan, MOST 108-2112-M-004-002-MY2.
P.-Y.C.  was supported by the Young Scholar Fellowship Program under MOST.
This work was supported by the MOST under grant No. 110-2636-M-007-007.
\end{acknowledgments}

\appendix

\section{Localization length}
\label{App:Loc}
When electrons are localized, the wave function exponentially decays with length, i.e. $\phi_L \propto e^{-L/\lambda}$, where $\phi_L=\sum_{n=1}^L(\phi_{na},\phi_{nb})^{T}c_n^{\dagger}$ is the eigenstate of the Hamiltonian $H=H_o+H_U$ with length $n$,  $\phi_{na/b}$ are the coefficients for sublattice $a/b$ at site $n$ and $\lambda$ is the localization length. The Schrodinger equation for zero eigenenergy state becomes
\begin{eqnarray}
(J_0+U_{1n})\phi_{nb}+(J_1+U_{2n})\phi_{n-1,b}=0,\\
(J_0+U_{1n})\phi_{na}+(J_1+U_{2n})\phi_{n+1,a}=0. 
\label{eq:tr}
\end{eqnarray}

The above equations give the ratio of coefficients between the first and the last site,  $\big|\phi_{La}\big|=\prod_{n=1}^L\big|\frac{J_1+U_{2n}}{J_0+U_{1n}}\phi_{1a}\big|$
	and 
	$\big|\phi_{Lb}\big|=\prod_{n=1}^L\big|\frac{J_0+U_{1n}}{J_1+U_{2n}}\phi_{1b}\big|$
	for each sublattice, respectively.
The final localization length for the system is the minimum of that of the sublattices. 
Thus, the localization length is given by
\begin{eqnarray}
\frac{1}{\lambda}=\frac{1}{L}\ln\prod_{n=1}^L\big| \frac{J_1+U_{2n}}{J_0+U_{1n}}\big|.
\end{eqnarray}
The equation can be solved analytically \cite{Shem2014}. 

Another approach to calculate the localization length is via Green's function.  
The localization length $\lambda$ is defined by
\begin{eqnarray}
\frac{2}{\lambda}=-\lim_{L\rightarrow \infty}\frac{1}{L}\mathrm{Tr}\ln |G_{1,L}|^2,
\label{eq:lambda}
\end{eqnarray}
where $n$ is the total number of sites of the one-dimensional Hamiltonian, $G_{1,L}$
is the propagator connecting the first and the last slice of the
system \cite{Mackinnon1983}. $G_{1,n}$ is computed with the iterative Green's function method \cite{Mackinnon1983,Kramer1993, Lewenkopf2013} by computing the onsite Green's function
$G_{n,n}=\left(E-h_n-U_fG_{n-1,n-1}U_b\right)$ and $G_{1,n}=G_{1,n-1}U_bG_{n,n}$
recursively till $n$ is large enough for convergence, where $h_n=(J_0+U_{1,n})\sigma_x$, $U_{f(b)}=(J_1+U_{2n})(\sigma_x+(-)i\sigma_y)/2$ and $U_{1(2)n}$ are defined in the main text.
 Within this method, the Hamiltonian is constructed in a slicing scheme, i.e. 
  \begin{eqnarray}
 H_{N}=\sum_{i=1}^{N}\left( |i\rangle h_i\langle i|+ |i\rangle U_b\langle i+1|+|i+1\rangle U_f\langle i|\right) 
 \end{eqnarray} 
 for the system with $N$ slices, 
  where $|i\rangle$ is the state for the $i$-th slice, $U_{f(b)}$ is the forward (backward) hopping matrices between the neighboring slices, and

 To calculate the Greens function for the system with $N+1$ slices, 
  the Hamiltonian for $N+1$ slices is
 \begin{eqnarray}
 	H_{N+1}=H_{N}+|N+1\rangle h_{N+1}\langle N+1|+H',
 \end{eqnarray}
 where $h_{N+1}$ is the Hamiltonian for $N+1$-th slices, the hopping matrix $H'=|N\rangle U_b\langle N+1|+|N+1\rangle U_f\langle N|$ between the $N-$th and $N+1-$th slice is treated as a perturbing term to $H_{N}+|N+1\rangle h_{N+1}\langle N+1|$.
 According to Dyson equation, the perturbed Greens function is given by $G_{N+1}=G_o+G_oH'G_{N+1}$, where $G_o=G_N+|N+1\rangle \left(E-h_{N+1}\right)^{-1} \langle N+1|$. Substitute $H'$ into the Dyson equation, one obtains the Greens function for $N+1$ slices ($G_{N+1}$)
 in which the submatrices are given by
 \begin{widetext}
 \begin{eqnarray}
 \langle N+1|G_{N+1}|N+1\rangle &=\left(E-h_{N+1}-U_f\langle N|G_{N}|N\rangle U_b\right)^{-1}, \label{eq:gnn}\\
 	\langle 1|G_{N+1}|N+1\rangle &= \langle 1|G_{N}|N\rangle U_b \langle N+1|G_{N+1}|N+1\rangle. \label{eq:g1n}	
 \end{eqnarray}
  \end{widetext}
 Eqs.~(\ref{eq:gnn}) and (\ref{eq:g1n}) are the main iterative equations for obtaining the localization length shown in Fig. \ref{meanwind1}(a).

\section{Symmetry analysis and topological classification}
\label{App:Symm}

The flattened Hamiltonian formalism allows us to construct  
 the effective Hamiltonian, 
 \begin{align}
 \mathcal{H}_{\rm eff} (k,t)=e^{-i\mathcal{H}^F_0t}\mathcal{H}_{\rm pre}(k)e^{i\mathcal{H}^F_0t}.
 \end{align}
 The topological invariants can be classified according to the symmetries of the effective Hamiltonian. 
 For the pre-quench Hamiltonian $\mathcal{H}_{\rm pre}(k)=-\sigma_z$ and the post-quench Hamiltonian $\mathcal{H}_0(k)=h_x(k)\sigma_x+h_y(k)\sigma_y$, one has the effective Hamiltonian 
 \begin{eqnarray}
 	\mathcal{H}_{\rm eff} (k,t) &=& \frac{h_y(k) \sin 2t}{\sqrt{h_x(k)^2+h_y(k)^2}} \sigma_x  \notag\\
	&&- \frac{h_x(k) \sin 2 t}{\sqrt{h_x(k)^2+h_y(k)^2}} \sigma_y 
 	+ \cos 2t \sigma_z.
 \end{eqnarray}
 The effective Hamiltonian breaks the particle-hole symmetry explicitly,
 but preserves the time-reversal symmetry $\mathcal{TH}_{\rm eff}(k, t )\mathcal{T}^{-1}=\mathcal{H}_{\rm eff}(-k,-t)$,
 and the additional two two-fold symmetries $\sigma_z\mathcal{H}_{\rm eff}(k, t )\sigma_z=\mathcal{H}_{\rm eff}(k,-t)$, 
 $\sigma_x\mathcal{H}_{\rm eff}(k, t )\sigma_x=-\mathcal{H}_{\rm eff}(-k,t)$.
 These two additional symmetries together with the time-reversal symmetry lead to a $\mathbb{Z}$ classification in $(1+1)$ dimensions.
The former two-fold symmetry acts like the reflection symmetry in the time domain. 
There are two fixed points $t=0$ and $\pi/2$ such that $[\sigma_z, \mathcal{H}_{\rm eff}(k,0)]=[\sigma_z, \mathcal{H}_{\rm eff}(k,\pi/2)]=0$.
The dynamical Chern number in this effective Hamiltonian is quantized in the half of the momentum-time space $k\in [0, 2\pi]$, $t\in [0,\pi/2]$  \cite{Chiu2013,Chiu2015,Morimoto2013,Shiozaki2014}.

The effective Hamiltonian has the following symmetries
\begin{align}
&\mathcal{T H}_{\rm eff}(k,t) \mathcal{T}^{-1} =\mathcal{H}_{\rm eff} (-k,-t), \notag\\ 
&\mathcal{R}_{t} {H}_{\rm eff}(k,t) \mathcal{R}^{-1}_t =\mathcal{H}_{\rm eff} (k,-t), \notag\\
&\mathcal{M}_{x} \mathcal{H}_{\rm eff}(k,t) \mathcal{M}_x^{-1} =-\mathcal{H}_{\rm eff} (-k,t),
\end{align}
where $\mathcal{T}^2 = \mathcal{R}_{t}^2=  \mathcal{M}_{x}^2= 1$, $\{ \mathcal{R}_{t},  \mathcal{M}_{x}\} =0$, and $[ \mathcal{T}, \mathcal{R}_{t}]=[ \mathcal{T}, \mathcal{M}_{x}]=0$. 

The  effective Hamiltonian can be expressed in terms of the effective massive Dirac Hamiltonian  $ \mathcal{H}_{\rm eff} (k,t) = k \gamma_1 + t \gamma_2 + M_0 \gamma_0$,
with $\{ \gamma_i, \gamma_j\} = 0$ ($i=0,1,2$).
We construct the minimal effective Dirac Hamiltonian
in terms of the tensor product form of the Pauli matrices
\begin{align}
&\gamma_1 = \sigma_x \otimes \sigma_x, \quad  
\gamma_2 = \sigma_y \otimes \mathbb{I}_{2\times2}, \quad
 \gamma_0 = \sigma_z \otimes \mathbb{I}_{2\times2},  \notag\\
& \mathcal{T} = \mathbb{I}_{2 \times 2} \otimes \sigma_z \mathcal{K}, \quad
 \mathcal{R}_t = \sigma_z\otimes \sigma_z, \quad
 \mathcal{M}_x = \sigma_x\otimes\mathbb{I}_{2\times2}.
\end{align}
One can check the only allowed mass term which preserving all the symmetries is the $ \gamma_0 $.
For the $\mathbb{Z}$ classification, we need to make copies of the original effective Hamiltonian. For simplicity, we just make one copy.
The double Hamiltonian is  $ \mathcal{H}_{\rm eff} (k,t) = k \gamma_1 \otimes  \mathbb{I}_{2\times2}  + t \gamma_2  \otimes  \mathbb{I}_{2\times2} + M_0 \gamma_0  \otimes  \mathbb{I}_{2\times2} $,
for which there are no other symmetry-preserving mass terms. This indicates that different phases are not adiabatically connected in this system.
On the other hand, we can flip one momentum of the copy and construct the double Hamiltonian,  
$ \mathcal{H}_{\rm eff} (k,t) = k \gamma_1 \otimes \sigma_z  + t \gamma_2  \otimes  \mathbb{I}_{2\times2} + M_0 \gamma_0  \otimes  \mathbb{I}_{2\times2} $.
There is another symmetry allowed mass term (anti-commute with $ \gamma_0  \otimes  \mathbb{I}_{2\times2}\otimes  \mathbb{I}_{2\times2}$), $M_1 = \sigma_y   \otimes  \sigma_y   \otimes  \sigma_y$. This indicates the systems are all in the same phase. 
We conclude from the above analysis that the system belongs to a $\mathbb{Z}$ classification.
Similar classification schemes can be found in Refs. [\onlinecite{Chiu2013,Chiu2015,Morimoto2013,Shiozaki2014}].

\section{Wannier center with disorders}
\label{App:Wannier}
In translational invariant systems, the Wannier orbits are constructed from the Bloch states $u_{n {\bf k}}({\bf r})$,
$w_n ({\bf r-R})= \frac{1}{\Omega} \int d {\bf k} e^{i {\bf k \cdot (r -R)}} u_{n {\bf k}}({\bf r}) $, with $\Omega$ being
the volume of the system, $\bf R$ is the position of the unit-cell, and $\bf r$ is the local position of the Wannier orbits within the unit-cell.
In an insulator, these Wannier orbits are localized states and are the eigenstates of the projected position operator $X_P = P X P$,
where $P$ is the projector to the occupied states which are well defined in an insulator.

To construct the Wannier orbits without using Bloch states, we first write down the Hamiltonian in the real space $\mathcal{H}_{IJ}$,
where $I(J)$ includes the band indices and positions. The spectrum can exhibit a gap and the corresponding occupied states $|\psi_{\alpha I}  \rangle$ are well defined. Here $\alpha$ is the eigen-energy index. The corresponding projectors are $P_{IJ}=\sum_{\alpha \in {\rm occ.} } |\psi_{\alpha I}  \rangle  \langle \psi_{\alpha J}  |$. The position operator can be defined by as a diagonal matrix ${\rm diag}(1,\cdots,1, 2, \cdots2, \cdots, N,\cdots, N)$, where $N$ is the total number of sites and at each site there are $L$ bands. The projected position operator can be constructed as usual $X_P=PXP$ \cite{Kivelson1982}.

\begin{figure}[]
	\includegraphics[width=0.5\textwidth]{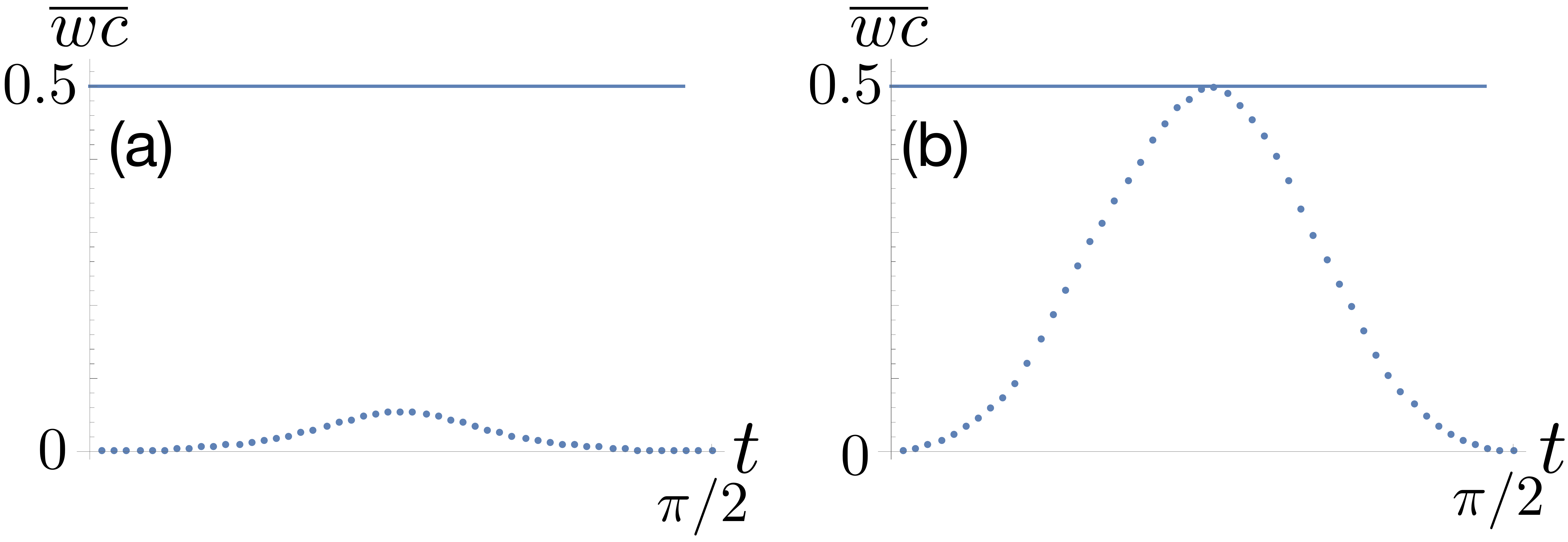}
	\caption{The Wannier center as a function of $t$.
	(a) Disorder-free  Hamiltonian $(\frac{J_0}{J_1},  \frac{W_0}{J_1})=(1.1, 0)$,  
	(b) disordered  Hamiltonian $(\frac{J_0}{J_1},  \frac{W_0}{J_1})=(1.1, 3)$. 
	There are $100$ disorder realizations.}
	\label{WC}
\end{figure}

Since the Wannier orbits are the eigenstates of the $X_P$, we can find diagonalize the $X_P$ and get the set of eigenstates.
If the set of the eigenstates are localized states, then these states are the Wannier orbits and the corresponding eigenvalues are the position of the Wannier states. The Wannier center of a localized state in $M$-th site can be defined  as $wc_M =|\langle w_M | X_P|w_M \rangle -M|$. We have $0<\langle x_M \rangle<1$. We can further define the average Wannier center $\overline {wc} =\frac{1}{N}\sum_{m=1}^N wc_M $.
In the presence of the chiral symmetry in one dimension gapped systems, the average Wannier center can have two values $\overline {wc}=0$ and $0.5$. 
The former corresponds to a trivial phase and the latter is the topological phase.
Although in the quench setup, the effective Hamiltonian does not have the chiral symmetry,
we observe the Wannier center of the topological post-quench state reaches $\overline {wc}=0.5 $ [Fig. \ref{WC}(b)].
On the other hand, for the trivial post-quench state, the Wannier center is below $0.5$ [Fig. \ref{WC}(a)].

\section{Correlation function formalism in quench setups}
\label{App:CA}
We consider an initial state contains $N$ particles. Each single-particle state we denote by $|\phi_\alpha({\bf x}) \rangle, \alpha = 1, \cdots, N$,
${\bf x}$ is the internal degrees of freedom, including position, spin, and the band. We require these single-particle states are orthonormal, 
$\sum_{\bf x} \langle \phi_\alpha ({\bf x})|\phi_\beta({\bf x}) \rangle=\delta_{\alpha,\beta}$. The N-particle initial state can be expressed as the Slater determinant of these single-particle state, 
\begin{align}
|\Psi_0\rangle = {\rm Det} [| \phi_i ({\bf x}_j) \rangle], \quad i,j=1,\cdots, N.
\end{align}
We consider an unitary evolution of this initial state $|\Psi_0\rangle$ by a static Hamiltonian $H = \sum_{\bf x, x'} \mathcal{H}_{\bf x, x'} c^\dagger _{\bf x}c_{\bf x'}$, where $c^{(\dagger)}_{\bf x}$ is the annihilation (creation) operator. Each single-particle state under this evolution is
$|\phi_\alpha({\bf x},t) \rangle = \sum_{\bf x'}  \exp [ - i \mathcal{H}_{\bf x, x'} t]|\phi_\alpha({\bf x}') \rangle, \alpha = 1, \cdots, N$.
The post-quench N-particle state is
\begin{align}
|\Psi(t)\rangle = e^{- i H t}|\Psi_0\rangle= {\rm Det} [| \phi_i ({\bf x}_j,t) \rangle]= \prod_i d_i^\dagger (t) |0 \rangle,
\end{align}
where 
\begin{align}
d_i^\dagger (t) &= e^{-i H t } d_i^ \dagger e^{i H t } =  e^{-i H t } \sum_{\bf y} V_{i{\bf y}} c^\dagger_{\bf y} e^{i H t }  \notag\\
&= \sum_{\bf x, y} V_{i {\bf y }} U_{\bf y,x}(t) c^\dagger_{\bf x}
\end{align}
with $U_{\bf y,x}(t)=e^{- i \mathcal{H}_{\bf y,x} t}$ and $V_{i{\bf y}} $
being an unitary matrix that rotates $d_i^ \dagger$ to $c_{\bf y}^\dagger$.

The post-quench single-particle state is 
\begin{align}
d_i^\dagger (t)|0\rangle =\sum_{\bf x, y} V_{i {\bf y }} U_{\bf y,x}(t) c^\dagger_{\bf x} | 0 \rangle 
= \sum_{\bf x} |\phi_i({\bf x},t) \rangle.
\end{align}

\begin{widetext}
The correlation function constructed from the N-particle post-quench state is
\begin{align}
C_{\bf x, x'}(t) &= \langle\Psi(t)| c^\dagger_{\bf x} c_{\bf x'} |\Psi(t) \rangle = \langle 0 | \prod_\alpha d_\alpha e^{i H t } c^\dagger_{\bf x} c_{\bf x'}  e^{-iHt}
\prod_\beta d^\dagger_\beta| 0\rangle\rangle  \notag\\
& =   \langle 0 |   \prod_\alpha d_\alpha [ \sum_{{\bf y},i} d_i  U_{\bf x, y}(t) V_{{\bf y} i}]^\dagger  [ \sum_{{\bf y'},j}  U_{\bf x', y'}(t) V_{{\bf y'} j} d_i ]
\prod_\beta d^\dagger_\beta|   | 0\rangle \notag\\
& = \sum_i [ \sum_{\bf y} U_{\bf x, y}(t) V_{{\bf y}i} ]^\dagger [ \sum_{\bf y'} U_{\bf x', y'}(t) V_{{\bf y'}i}  ] \notag\\
&= \sum_i | \phi_i({\bf x'},t)  \rangle \langle \phi_i({\bf x},t)  |.
\end{align}
\end{widetext}

The correlation matrix can be used for computing the entanglement spectrum. 
The existence of the crossings in the entanglement spectrum can detect the topology of the post-quench state as demonstrated in several examples.

\section{Other parameters for the disorder-induced topology in quench dynamics}
\label{App:other}
\begin{figure}[]
	\includegraphics[width=0.35\textwidth]{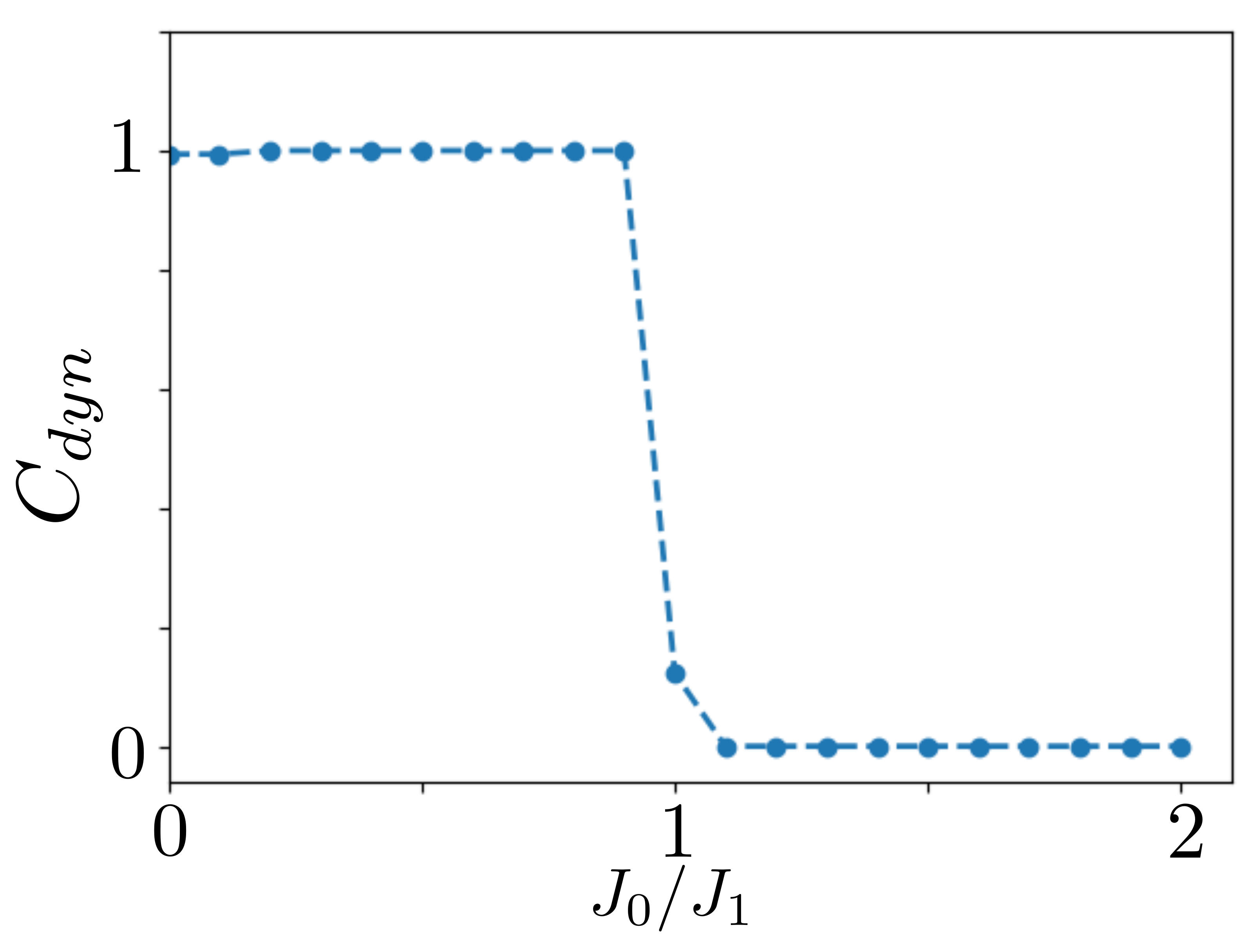}
	\caption{The dynamical Chern number (DCN) for the static Hamiltonian in the clean limit. The parameters are $J_0=1.1, J_1=1, N_x=400, L =20$.
	Here $L$ is the number of mesh points for the twisted boundary condition.}
	\label{fig:cleandcn}
\end{figure}

 In the clean limit, the dynamical Chern numbers are calculated for $0\leq J_0\leq 2$ and $J_1=1$ of the SSH Hamiltonian $H_o$. The results are shown in Fig. \ref{fig:cleandcn}. For $J_0>1$, the static Hamiltonian becomes trivial and the dynamical Chern number is zero. 

\begin{figure}[]
	\includegraphics[width=0.5\textwidth]{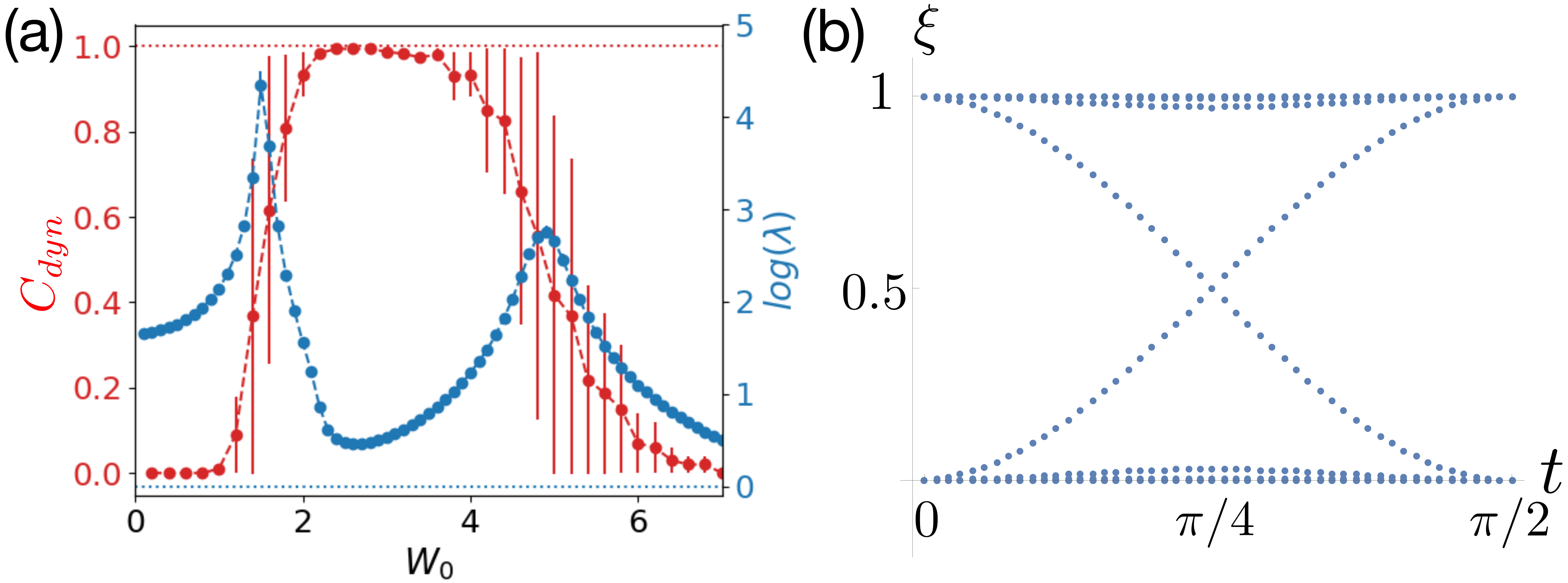}
	\caption{(color online) (a) The disorder averaged mean dynamical Chern number and localization length obtained from Eq.~(\ref{eq:lambda}) for the quench Hamiltonian. The error bar is the standard deviation. The parameters are $J_0=1.1, J_1=1, N_x=100, L=400$. (b) The entanglement spectrum of the post-quench state with $W_1=3$.
 The parameters are $J_0=1.1, J_1=1, N_x=400$. There are more than $50$ disorder realizations for each data point. }
	\label{meanwind1}
\end{figure}

We consider the case with vanishing intercell disorder $W_2=0$.
We find for $2.2 \lesssim W_1 \lesssim 3.8$, the dynamical Chern number is close to an integer with vanishing fluctuations as shown in Fig. \ref{meanwind1}(a).
The phase boundaries, where the  dynamical Chern number is close to half-integer, are at $W_0=1.5, 4.9$. The localization length $\lambda$ also indicates  
delocalized transitions at the same values of $W_0$ [Fig. \ref{meanwind1} (a)].
The entanglement spectrum has a crossing at $t=\pi/4$ when the post-quench state has integer dynamical Chern number $W_1=3$ [Fig. \ref{meanwind1}(b)].

\bibliography{TI-references,references}

\end{document}